# Change in flexibility of DNA with binding ligands

Anurag Singh and Amar Nath Gupta*
Soft Matter Laboratory, Department of Physics, Indian Institute of Technology Kharagpur, India-721302
E-mail: amar1518@gmail.com ; ang@phy.iitkgp.ernet.in Phone: +91 3222-283810

**Abstract**

The percentage and sequence of AT and GC base pairs and charges on the DNA backbone contribute significantly to the stiffness of DNA. This elastic property of DNA also changes with small interacting ligands. The single-molecule force spectroscopy technique shows different interaction modes by measuring the mechanical properties of DNA bound with small ligands. When a ds-DNA molecule is overstretched in the presence of ligands, it undergoes a co-operative structural transition based on the externally applied force, the mode of binding of the ligands, the binding constant of the ligands to the DNA, the concentration of the ligands and the ionic strength of the supporting medium. This leads to the changes in the regions- upto 60 pN, cooperative structural transition region and the overstretched region, compared to that of the FEC in the absence of any binding ligand. The cooperative structural transitions were studied by the extended and twistable worm-like chain model. Here we have depicted these changes in persistence length and the elastic modulus constant as a function of binding constant and the concentration of the bound ligands, which vary with time. Therefore, besides ionic strength, interacting proteins and content of AT and GC base pairs, the ligand binding or intercalation with the ligands is an important parameter which changes the stiffness of DNA.

**Keywords**: ds-DNA, FEC, Single molecule force spectroscopy, Persistence length, elastic modulus, ligand, flexibility

## Introduction

In several biological processes, a double stranded DNA is converted into single stranded DNA. The dynamics of such process is an open question. In DNA replication [1], such conversion enables it to duplicate itself during cell division. When a cell prepares to divide, the DNA helix splits down the middle and becomes two single strands. These single strands serve as templates for building new complimentary strands. Thus, forming two double-stranded DNA molecules each a replica of the original DNA molecule. Similarly, during DNA transcription, [1] RNA polymerase creates a transcription bubble which separates the two strands of the DNA helix. This is done by breaking the hydrogen bonds between complementary DNA nucleotides. The stiffness of the DNA affects the replication and transcription processes.

In DNA-protein or DNA-ligands interactions, ligands may bind to ds-DNA through different binding modes. They may bind into major or minor grooves of the double helix, intercalate between bases or associate with melted single strand of DNA. It was shown using optical tweezers that the flexibility of DNA is enhanced when bound with protein like HMGB [2] where the persistence length changed from 50 nm to 10 nm approximately in the presence of NaCl. The flexibility of DNA depends on its base sequence: GC base-paired regions are expected to be stiffer than AT regions as GC base pairs are more strongly bound.

In past, single molecule force spectroscopy has been extensively used to study the well characterized structural transitions of a DNA molecule. When a DNA molecule is stretched between AFM tip and the substrate in the absence of any binding agent, the change in mechanical properties of the molecule depends on the strength of the applied force. This change in the mechanical properties of DNA is depicted by its force-extension characteristic curve. DNA in a solution medium adopts random coil conformation which corresponds to maximum entropy. The process of pulling of the DNA molecule reduces this entropy and that costs energy. The entropic force is generally weak and is less than 10 pN. After this force regime, the enthalpic force takes over and DNA stretches like a spring. Up to 50 pN force regime, the curve can be described by standard worm-like chain model [3]. The molecule undergoes a cooperative structural transition at around 60 pN which is reflected as plateau region in the force extension characteristic curve (FEC). Recent studies attribute this plateau region to the phenomenon of DNA melting [4-5]. DNA melting is the result of breakage of hydrogen bonds and rotation of DNA strands in order to reduce the torsional strain during the overstretching transition [6]. Further increment in the force results into another co-operative transition around 125 pN and if the force is further increased, it results in the separation of ds-DNA into two single stranded DNA [7]. Recently, studies have also shown that the structural plasticity of stretched DNA also depends on its torsional constraint state [8-9]. In our study, we have considered torsionally unconstrained ds-DNA.

If ds-DNA molecule is stretched in the presence of binding ligands then their different binding modes are characterized by different FECs. In past, these FECs have been obtained for the cross-linking anti-cancer drug: cisplatin, groove binder: berenil and the intercalating dye: ethidium bromide [10]. The FECs were measured with time at a particular concentration of the drug in the supporting medium. In this study, we report the dependence of the stiffness of DNA on the concentration of ligands bound to the DNA. The ligand concentration plays an important role in the conversion of stiff rod-like structure to a flexible molecule. We have used the extended and twistable worm-like chain model to describe how persistence length and elastic modulus change with ligand concentration.

**Methods and Results**

The over-stretching saw-tooth like pattern was observed experimentally at the single-molecule level when tension was applied to ds-DNA. At high force, this leads to generation of ss-DNA which can be explained by twisted worm-like-chain model. The force dependent elasticity of the ds-DNA can be explained by introducing a complex term consisting of twist rigidity and stretching modulus. In this study, we propose a model to describe the effect of mechanical stress in the presence of the binding ligands. The change in the mechanical properties of the over-stretching DNA is a continuous process. The concentration of ligand/drug binding to the DNA continuously increases until it saturates to the final concentration of the binding agent in the buffer solution. In this study, we have developed a minimal model which describes the dependence of persistence length and the stiffness constant of the DNA on the binding constant and the concentration of the binding ligands/drugs. The experimental data of stretching of single-molecule of ds-DNA shows over-stretching at two different forces. First at around 65 pN where there is a change in the extension at constant force and second at around 125 pN where there is a generation of ss-DNA. We have used a master equation to capture these two different transitions.

$$x(F) = \left[ L_{c0} \left( 1 - \frac{1}{2} \sqrt{\left( \frac{kT}{F L_P} \right)} + \frac{F}{K} \right) + \frac{n_1 \Delta x_1}{1 + e^{(F_{h1} - F)\Delta x_1 / kT}} + \frac{n_2 \Delta x_2}{1 + e^{(F_{h2} - F)\Delta x_2 / kT}} \right] \text{---------} \quad (1)$$

The first term in eqn.(1) represents the worm-like chain (WLC) model, the second term represents the co-operative structural transition region which is obtained after the first transition at 65 pN and the third term represents the region of the FEC which is obtained after the second transition at around 125 pN. The combination of all the three extensions gives the total extension of the DNA corresponding to the force applied at that instant [11-15].

The FECs of ds-DNA in the presence of two different ligands at various concentrations have been fitted with eqn.(1) where; $L_{c0}$ is the contour length of the ds-DNA, k is Boltzmann constant, T is the temperature, F is the applied force, $L_P$ is the persistence length, K is the elastic modulus constant, $n_1$ and $n_2$ are number of nucleotides unfolding during the transitions, $\Delta x_1$ and $\Delta x_2$ are the extension changes upon unfolding at force F corresponding to first ($F_{h1} = 65$ pN) and second ($F_{h2} = 125$ pN) transitions respectively. There are many binding pockets available along the DNA strand where small ligands, protein molecules can bind and modulate the elastic nature of the helix. The stiffness of the helix changes with the concentration of the binding ligands. So, the $L_P$ of DNA changes with pH, ionic strength of the supporting medium [16] and many other parameters. We have incorporated the dependence of $L_P$ and K on the ligand concentration in eqn.(1) to capture the effect of bound ligands to the DNA by following relations;

$$L_p(x) = (50 + \gamma) - \beta e^{[\tanh(x-\alpha)]} \quad \quad \quad \quad \quad \quad \quad \quad (2)$$

$$K(x) = \gamma' + \beta' e^{[\tanh(x-\alpha')]} \quad \quad \quad \quad \quad \quad \quad \quad (3)$$

The empirical relation consists of six parameters. The parameters α and α' are related to the critical concentration of the binding ligands. After this critical concentration, there is an appreciable change in $L_P$ and K respectively. This is reflected as change in the FEC. The parameters β and β' allows the function to take all possible values of $L_P$ and K respectively. The parameters $\gamma$ and $\gamma'$ are the offset values for $L_P$ and K respectively. In the above equations, *x* represents the concentration of the ligands which varies with time as it binds to the DNA molecule. The binding of interacting ligands to DNA molecule can be given by change in the signal amplitude obtained from surface plasmon resonance (SPR) measurements [17],

$$\Delta R(t)_{bind} \propto e^{(1-\tau t)} \quad \quad \quad \quad \quad \quad \quad \quad (4)$$

where $\tau = k_a x(t) + k_d$, which represents the rate of increase of the signal amplitude and x is the ligand concentration. The association ($k_a$) and dissociation ($k_d$) of ligands can be estimated by

$$K_D = k_d/k_a = 1/B \quad \quad \quad \quad \quad \quad \quad \quad (5)$$

where B is the binding constant of the ligand bound to the DNA. The binding constant for cisplatin, B (cisplatin) = $5.73 (\pm 0.45) \times 10^4$ /M , [18] was obtained using Fourier transform infrared, ultraviolet-visible, and circular dichroism spectroscopic methods. Spectroscopic evidence showed that cisplatin binds to the guanine N7 site with minor perturbations of the backbone phosphate group. The binding constant for berenil, B (Berenil) = $6.8(\pm 0.25) \times 10^3$ /M [19] was obtained using Scatchard analysis. This shows that cisplatin is relatively better binder than berenil to the DNA.

The results obtained after fitting the data with eqn.(1) are summarized in Fig. 1 and fitting parameters are tabulated in Table 1.

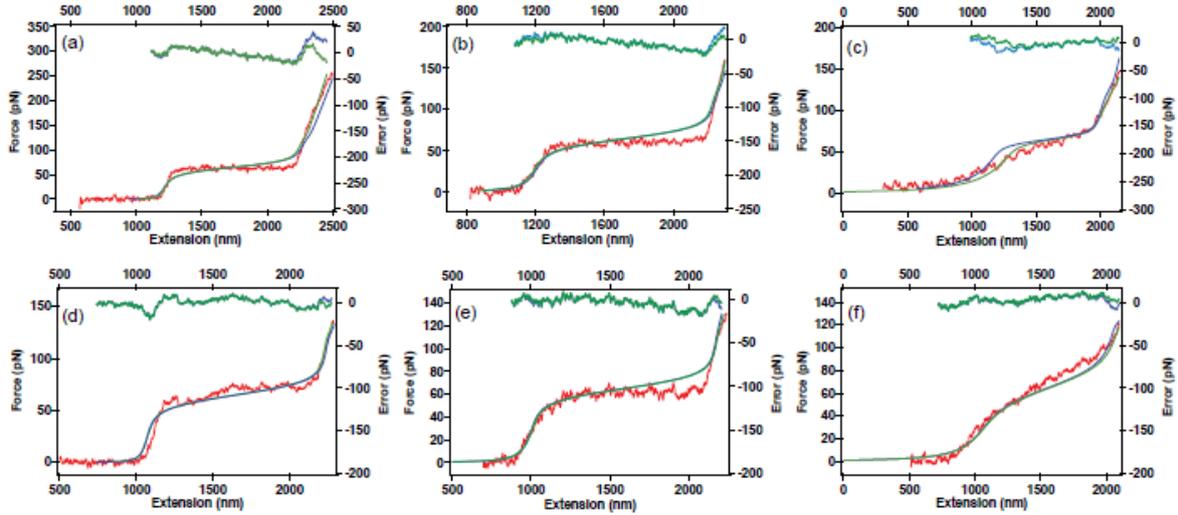

Figure 1: (color online) The figure represents the results of the fitting of the experimental data (red) using master equation (blue) and using modified master equation (green). The subplot in the figures gives the deviation of the fitted curve from the experimental data. Fig 1(a), (b) and (c) gives the results obtained for cisplatin at 0 hr, 1 hr and 24 hr respectively. Fig 1(d), (e) and (f) gives the results obtained for berenil at 0 µg/ml, 1.5 µg/ml and 15 µg/ml respectively.

The prime effect of the ligand attaching to the ds-DNA, irrespective of its binding mode, is to increase the flexibility of ds-DNA. The observation of the fitting parameters namely $L_{c0}$, $L_P$, and K indicates that the increase in the flexibility of ds-DNA is accompanied by the increase in $L_{c0}$, decrease in $L_P$, and increase in K of the ds-DNA measured at fixed temperature. This indicates the conversion of ds-DNA into ss-DNA when force is applied on the ds-DNA in the presence of the binding ligand. The parameters $n_1$ and $n_2$ are the number of base pairs coming apart in the co-operative structural transition regions. We have observed that even after 1 hr (in case of cisplatin) and at 1.5 µg/ml (in case of berenil), there is only a small change in the values of $n_1$ and $n_2$. This indicates that only a small amount of ligand has bound to the ds-DNA. At 24 hr (in case of cisplatin) and at 15 µg/ml (in case of berenil), there is an appreciable decrease in the value of $n_1$ and increase in the value of $n_2$, see Table 1.

Table 1: The FECs were fitted by eqn.(1) and the fitting parameters are listed here ($kT=4.06$ pN-nm)

| Parameters | Cisplatin Time(hr) | | | Berenil Conc. (µg/ml) | | |
|---|---|---|---|---|---|---|
| | 0 | 1 | 24 | 0 | 1.5 | 15 |
| $L_{c0}$ (nm) | 1220±15 | 1270±15 | 1450±15 | 1085±15 | 1050±15 | 1250±15 |
| $L_p$ (nm) | 40±4 | 7±1 | 0.5±0.01 | 35±4 | 7±1 | 1±0.02 |
| K (pN) | 860±90 | 1500±150 | 1800±190 | 1100±120 | 1250±130 | 1500±160 |
| $n_1$ | 915±10 | 920±10 | 680±7 | 1070±11 | 1070±11 | 830±9 |
| $L_{c1}$ ($\Delta x_1$) | 0.6±0.01 | 0.62±0.01 | 1.05±0.02 | 0.64±0.01 | 0.7±0.01 | 0.4±0.01 |
| $n_2$ | 30±3 | 25±3 | 50±5 | 30±3 | 20±2 | 90±9 |
| $L_{c2}$ ($\Delta x_2$) | 0.6±0.01 | 0.6±0.01 | 0.6±0.01 | 0.7±0.01 | 0.7±0.01 | 1.2±0.03 |

This suggests that there is a reduction in the co-operative structural transition region at 65 pN and increase in co-operative structural phenomenon at higher forces. The parameters $L_{c1}$ and $L_{c2}$ are relative contour length in the cooperative structural transition regions. The values of $L_{c1}$ and $L_{c2}$ are increasing with the increment in time or concentration. This supports the increased flexibility of ds-DNA with increment in the bound ligand concentration. After including the dependence of $L_P$ and K on concentration of the binding ligand in the eqn.(1), we have again fitted the experimental data and the fitting parameters are tabulated in Table 2.

Table 2: The FECs were fitted with modified eqn.(1) and the additional fitting parameters are listed here

| Parameters | Cisplatin Time(hr) | | | Berenil Conc. ($\mu g/ml$) | | |
|---|---|---|---|---|---|---|
| | 0 | 1 | 24 | 0 | 1.5 | 15 |
| $\alpha$ | 3±0.5 | 3±0.5 | 3±0.5 | 2.5±0.5 | 2.5±0.5 | 3.5±0.5 |
| $\beta$ | 30±3 | 29±3 | 30±3 | 22±2 | 21±2 | 20±2 |
| $\gamma$ | 34±4 | 32±4 | 32±4 | 7±1 | 7±1 | 1.5±0.1 |
| $\alpha'$ | 4±0.5 | 4±0.5 | 4±0.5 | 3.5±0.5 | 2.0±0.5 | 3.6±0.5 |
| $\beta'$ | 1180±120 | 700±70 | 150±15 | 2000±200 | 450±45 | 620±60 |
| $\gamma'$ | 535±50 | 545±50 | 530±50 | 500±50 | 150±15 | 160±15 |

We have observed that the constants α and α' are fixed for cisplatin but vary non-uniformly in case of berenil. This indicates that the critical concentration for cisplatin remains same at all values of time but it varies for berenil which can be seen by different critical concentration values for different concentrations of berenil. This may be primarily because of different binding mode through which these ligands bind to the ds-DNA. Another constant β, which corresponds to $L_p$, remains same for both the ligands but β', which corresponds to K, differs with time or concentration for both the ligands. The parameters γ and γ' are the offset values for $L_P$ and K respectively.

The FEC can be described by WLC model which is applicable at lower force regime up to 30 pN [20]. The FECs were fitted with WLC model with parameters $L_{c0}$, $L_P$, and K. The same parameters were used to fit the experimental data with eqn.(1) in the low force regime. The obtained values of fitting parameters are presented in Table 3.

Table 3: The worm-like chain model was applied in low force regime of FECs before the first transition and the fitting parameters are listed here ($kT=4.06$ pN-nm)

| Parameters | Cisplatin Time(hr) | | | Berenil Conc. ($\mu g/ml$) | | |
|---|---|---|---|---|---|---|
| | 0 | 1 | 24 | 0 | 1.5 | 15 |
| $L_{c0}$ (nm) | 1220±15 | 1270±15 | 1450±15 | 1085±10 | 1050±10 | 1250±15 |
| $L_P$ (nm) | 38±4 | 7±1 | 0.5±0.01 | 35±4 | 7±1 | 1±0.02 |
| K (pN) | 860±90 | 1500±150 | 1800±190 | 1100±120 | 1250±130 | 1500±160 |

We have further used the freely jointed chain (FJC) model to fit the FEC in higher force regime and the obtained parameters are listed in Table 4.

Table 4: The high force regime of FECs were fitted by the freely jointed chain model which behave like single stranded DNA ($kT=4.06$ pN-nm)

| Parameters | Cisplatin Time(hr) | | | Berenil Conc. ($\mu g/ml$) | | |
|---|---|---|---|---|---|---|
| | 0 | 1 | 24 | 0 | 1.5 | 15 |
| $L_{c0ss}$ (nm) | 2160±30 | 2160±30 | 2100±30 | 2020±20 | 2020±20 | 2000±5 |
| $L_{Pss}$ (nm) | 1.4±0.2 | 1.4±0.2 | 0.5±0.05 | 1.4±0.2 | 1.4±0.2 | 0.5±0.05 |
| $K_{ss}$ (pN) | 1800±200 | 2000±200 | 2000±200 | 900±90 | 1200±120 | 1250±125 |

These values of parameters correspond to single stranded DNA. The results of the fitting of the experimental data with WLC and FJC models are summarized in Fig. 2.

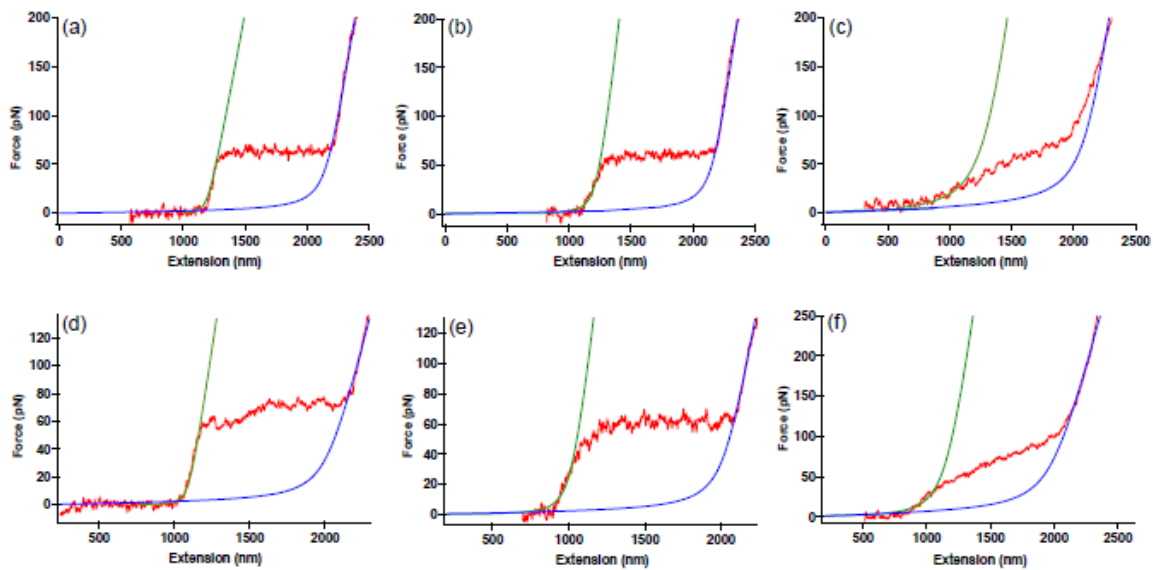

Figure 2: (color online) The figure represents the results of the fitting of the experimental data (red) using WLC model (green) and the FCJ model (blue). Fig 2(a), (b) and (c) gives the results obtained for cisplatin at 0 hr, 1 hr and 24 hr respectively. Fig 2(d), (e) and (f) gives the results obtained for berenil at 0 µg/ml, 1.5 µg/ml and 15 µg/ml respectively.

The number of base pairs of ds-DNA was estimated from the fitted value of contour length for cisplatin (#6260) and berenil (#5880) binders by dividing by 0.34 nm, consecutive base pair distance. The stability of ligand bound DNA was estimated by calculating the change in free energy per base pair at various time or concentration of ligand used in the experiment, presented in Table 5.

Table 5: The change in the free energy per base pair ($\Delta G$ in kJ/mol/bp) and the separation between consecutive base pairs $(\Delta x_1)_{bp}$ in force range 60-70 pN; $(\Delta x_2)_{bp}$ in the force range above 110 pN after the ligands are attached to the dsDNA has been estimated by integrating the area under the FECs in different regime.

| Parameters | Cisplatin Time(hr) | | | Berenil Conc. ($\mu$g/ml) | | |
|---|---|---|---|---|---|---|
| | 0 | 1 | 24 | 0 | 1.5 | 15 |
| $\Delta G$ | 4.6±0.5 | 4.5±0.5 | 4.2±0.5 | 5.4±0.5 | 5.3±0.5 | 4.7±0.5 |
| $(\Delta x_1)_{bp}$ | 0.5±0.05 | 0.52±0.05 | 0.53±0.05 | 0.52±0.05 | 0.51±0.05 | 0.52±0.05 |
| $(\Delta x_2)_{bp}$ | 0.68±0.08 | 0.68±0.08 | 0.65±0.08 | 0.71±0.09 | 0.71±0.09 | 0.69±0.09 |

The change in free energy per base pair decreases with increase in the number of bound ligand molecules. This shows the gain in the stability of DNA molecule. The distance between consecutive base pairs in the co-operative structural transition region (60-70 pN) and in the overstretched region (above 110 pN) was estimated by using the facts that the length of the ds-DNA increases by 70 % in co-operative structural transition region and overall length of the DNA increases approximately by 1.7 times its contour length. Recently, the transition mechanism of DNA over-stretching has been studied using molecular dynamics where the range of $(\Delta x)_{bp}$, in the absence of any ligand, has been proposed to be 0.34 nm-0.58 nm [21]. Also in a previous study, the distance between consecutive base pairs of ds-DNA complexed with ethidium bromide was reported to be 0.68 nm [10]. We have obtained the average $(\Delta x)_{bp}$ of DNA molecule in the presence of cisplatin as 0.52 nm in 60-70 pN force range and 0.68 nm above 110 pN force; and in the presence of berenil, it is 0.52 nm in 60-70 pN force range and 0.7 nm above 110 pN force. These observations show that both cisplatin and berenil are equally responsible for changing the flexibility of ds-DNA. These observations verify our proposed empirical relations for persistence length and the elastic modulus constant. Moreover, there was appreciable reduction in the errors after modification of eqn.(1) as shown in the Fig. 1. Thus, these empirical relations may be used to get better results to explore the mechanical properties of DNA.

**Discussion**

Many theoretical models like FJC, WLC have been used to analyse the elastic behaviour of a single molecule under tension. The elastic property of the DNA changes by bending, twisting and stretching of DNA molecule in various biological processes. Elastic nature of DNA depends on various factors including helical structure, content and sequence of AT and GC base pair, pH, temperature and ionic strength of surrounding medium. Recently, twistable WLC model [22-24] has been proposed to capture the DNA elasticity in the intermediate range of stretching force in pulling measurement by using optical tweezers [25]. In single-molecule pulling experiment, a known force is applied on a molecule via a trapped beads and corresponding extension is measured. The DNA over-stretching is a co-operative phenomenon where DNA gets changed in helical structure accompanied by the loss of hydrogen bonding.

The FEC of ds-DNA in the presence of different ligands were experimentally measured using AFM technique. The FECs were analysed by known parameters like $L_c$, $L_p$ and K. These parameters corresponds to ds-DNA and are in the expected range, see Table 1. The FEC of a single-molecule has a unique characteristic feature of providing information about the K of a single-molecule through the curvature of the FEC and $L_p$. The empirical relations were proposed for the variation of $L_p$ and K of DNA molecule with the bound ligand concentration, given by eqn.(2) and (3). The FECs can be reproduced for any of these parameter values which allow us to have a broader picture of change in the mechanical properties of the DNA while it is overstretched. The variation in $L_P$ and K of DNA with bound ligand is continuous rather being a constant value during the process of over-stretching of ds-DNA. In another experiment, the elastic parameters of DNA were measured at pH=7.0 for a fixed concentration of spermidine (100 μM) in $NaHPO_4$ solution. The change in $L_P$ was observed from 47.4 ± 1.0 nm to 38.7 ± 1.0 nm and change in K was observed from 1008 ± 38 pN to 1202 ± 83 pN [26]. This observation clearly shows that the small interacting ligand plays an important role in changing the elastic property of DNA.

In our analysis, the variation of $L_P$ from 50.0 ± 0.5 nm to 1.0 ± 0.5 nm and the variation of K from 400 ± 20 pN to 2000 ± 20 pN were captured by the proposed empirical relations. Thus, we can study all possible ranges of concentration of ligands which can positively bind to the DNA. While fitting the experimental FECs using eqn.(1), we have observed that the variation of $L_P$ was from 38.0 ± 1.5 nm to 1.5 ± 1 nm and the variation of K was from 400 ± 20 pN to 1800 ± 20 pN with time or various ligand concentrations. This continuous nature of $L_P$ and K were not studied against the binding ligand concentration in the past. We have presented a simple empirical relation which can predict the variation in $L_P$ and K with ligand concentration which is very difficult to estimate experimentally.

The continuous change in the average base pair distance was also estimated in the analysis of FECs, the obtained variation is listed in Table 5. The increment in $(\Delta x)_{bp}$ suggests how ds-DNA behaves like a ss-DNA in presence of the bound ligands. These ligands bound to DNA molecule change a rod like stiff ds-DNA to a flexible ss-DNA molecule.

**Conclusions**

Through single-molecule pulling experiments using AFM on complexes of ds-DNA and ligands, cooperative and structural transitions at various force regime were observed. The contour length of ds-DNA changes depending upon the strength of the force applied. In the beginning, where force is low, the change in contour length is not appreciable but as the force is further increased, 70% increase in the contour length is reported. Even in the presence of ligands, the contour length of ds-DNA increased as the applied force is increased. A master eqn.(1) along with empirical relations eqn.(2) and eqn.(3), depicting the continuous variation in the values of $L_P$ and K while stretching the complexes of ds-DNA and small bound ligands, has been proposed to explore the elastic behaviour of the biopolymer under tension.

The modification implemented in eqn.(1) has an advantage to address the changes in the mechanical properties of the DNA in the presence of bound ligands. Many enzymatic reaction and gene regulation processes depend on the thermal fluctuations of the shape of the DNA which is indirectly connected with the flexibility of the DNA molecule. The interactions between protein and DNA control many paramount cellular processes such as transcription, replication, DNA repair, recombination and other critical steps in cellular development [27]. The knowledge of elastic properties of DNA can be used to design DNA origami [28]. Small binding molecules are capable of blocking or mimicking these processes. Therefore, offering themselves as a potential pharmacological agent. The equilibrium binding constant depends on the concentration of both the protein and the DNA when protein binds to the specific site of the DNA. The kinetics of the interacting ligand depends on the binding constant. Thus, it is very important to have a better understanding of how the protein binding to the DNA changes the mechanical properties of the DNA. Moreover, in case of drug delivery via DNA, the elastic property of DNA plays an important role in determining the rate at which the specific drug can be administered in the affected area. These facts are important for designing a drug for a specific pharmacological function [29- 30].

There is future scope to explore the mixture of ligands interacting with the ds-DNA which mimics the cellular environment where different kinds of ligands can interact with the DNA at the same time. In such case, the effective change in the flexibility of ds-DNA is expected to be competitive depending on the binding constants of the interacting ligands. Moreover, a database can be created based on the binding constant of the ligands to ds-DNA which can be used to predict the mechanism of binding of unknown ligands to DNA. A very similar approach is used in the genetic screening where the assessment of an individual's genetic makeup is done to detect inheritable defects that may be transmitted to the next generation [31].

**Acknowledgement**

The authors thank Dr. Rupert Krautbauer of Ludwig-Maximilians Universität, Amalienstr Munich, Germany for sharing the single-molecule AFM data. We acknowledge Debajyoti De and Suparna Khatun for critical comments and productive discussions about the manuscript. AS acknowledges the support of institute fellowship and ANG acknowledges the support of ISIRD grant from the Indian Institute of Technology Kharagpur, India.

**References**
(1) Falkenberg, M.; Larsson, N.-G.; Gustafsson, C. M. Annu. Rev. Biochem. 2007, 76, 679–699.
(2) Zhang, J.; McCauley, M. J.; Maher, L. J.; Williams, M. C.; Israeloff, N. Nucleic Acids Res. 2009, 37, 1107–1114.
(3) Bustamante, C.; Smith, S. B.; Liphardt, J.; Smith, D. Curr. Opin. Struct. Biol. 2000, 10, 279–285.
(4) Rouzina, I.; Bloomfield, V. A. Biophys. J. 2001, 80, 882–893.
(5) Rouzina, I.; Bloomfield, V. A. Biophys. J. 2001, 80, 894–900.
(6) Storm, C.; Nelson, P. Phys. Rev. E 2003, 67, 051906.
(7) Rief, M.; Clausen-Schaumann, H.; Gaub, H. E. Nat. Struct. & Mol. Biol. 1999, 6, 346–349.
(8) King, G. A.; Peterman, E. J.; Wuite, G. J. Nat. Commun. 2016, 7, 11810.
(9) Sarkar, A.; Leger, J.-F.; Chatenay, D.; Marko, J. F. Phys. Rev. E 2001, 63, 051903.
(10) Krautbauer, R.; Pope, L. H.; Schrader, T. E.; Allen, S.; Gaub, H. E. FEBS letters 2002, 510, 154–158.
(11) Cecconi, C.; Shank, E. A.; Dahlquist, F. W.; Marqusee, S.; Bustamante, C. Eur. Biophys. J. 2008, 37, 729–738.
(12) Kosikov, K. M.; Gorin, A. A.; Zhurkin, V. B.; Olson, W. K. J. Mol. Biol. 1999, 289, 1301–1326.
(13) Kim, J.; Zhang, C.-Z.; Zhang, X.; Springer, T. A. Nature 2010, 466, 992–995.
(14) Pauling, L.; Corey, R. B. Proc. Natl. Acad. Sci. U.S.A. 1951, 37, 251–256.
(15) Tinoco Jr, I.; Bustamante, C. Biophys. Chem. 2002, 101, 513–533.
(16) Shokri, L.; McCauley, M. J.; Rouzina, I.; Williams, M. C. Biophys. J. 2008, 95, 1248–1255.
(17) Dee, D. R.; Gupta, A. N.; Anikovskiy, M.; Sosova, I.; Grandi, E.; Rivera, L.; Vincent, A.; Brigley, A. M.; Petersen, N. O.; Woodside, M. T. Biochimica et Biophysica Acta 2012, 1824, 826–832.
(18) N'soukpoé-Kossi, C. N.; Descôteaux, C.; Asselin, É.; Tajmir-Riahi, H.-A.; Bérubé, G. DNA Cell Biol. 2008, 27, 101–107.
(19) Bielawski, K.; Bielawska, A.; Słodownik, T.; Popławska, B.; Bołkun-Skórnicka, U. Acta Pol Pharm. 2008, 65, 135–140.
(20) Bouchiat, C.; Wang, M.; Allemand, J.-F.; Strick, T.; Block, S.; Croquette, V. Biophys. J. 1999, 76, 409–413.
(21) Bongini, L.; Lombardi, V.; Bianco, P. J. R. Soc. Interface 2014, 11, 20140399.
(22) Solanki, A.; Neupane, K.; Woodside, M. T. Phys. Rev. Lett. 2014, 112, 158103.
(23) Smith, S. B.; Cui, Y.; Bustamente, C. Science 1996, 271, 795.
(24) Strick, T.; Allemand, J.-F.; Croquette, V.; Bensimon, D. Prog. Biophys. Mol. Biol. 2000, 74, 115–140.
(25) Ritchie, D. B.; Woodside, M. T. Curr. Opin. Struct. Biol. 2015, 34, 43–51.
(26) Wang, M. D.; Yin, H.; Landick, R.; Gelles, J.; Block, S. M. Biophys. J. 1997, 72, 1335.
(27) Luscombe, N. M.; Austin, S. E.; Berman, H. M.; Thornton, J. M. Genome Biol. 2000, 1, 1.
(28) Rothemund, P. W. K. Nature 2006, 440, 297–302.
(29) Kerwin, S. M. Curr. Pharm. Des. 2000, 6, 441–471.
(30) Fujita, T.; Fujii, H. Int. J. Mol. Sci. 2015, 16, 23143–23164. 15
(31) Oberle, I.; Camerino, G.; Heilig, R.; Grunebaum, L.; Cazenave, J.-P.; Crapanzano, C.; Mannucci, P. M.; Mandel, J.-L. N. Engl. J. Med. 1985, 312, 682–686.